\documentclass[aps,prb,superscriptaddress,onecolumn,nofootinbib, noeprint, 11pt]{revtex4-2}

\usepackage[a4paper, top=0.75in, left=1in, right=0.9in, bottom=0.75in, includehead, includefoot]{geometry}
\usepackage{amsmath}
\usepackage{amssymb}
\usepackage[utf8]{inputenc}
\usepackage{jabbrv}
\usepackage{natbib}
\setcitestyle{numbers,sort&compress}
\usepackage{graphicx}
\usepackage{dsfont}
\usepackage{stmaryrd}
\usepackage{orcidlink}

\DeclareMathAlphabet{\pazocal}{OMS}{zplm}{m}{n}
\DeclareUnicodeCharacter{2212}{-}
\usepackage{booktabs}
\usepackage[english]{babel}
\usepackage{url}
\usepackage{hyperref}
\hypersetup{
  colorlinks=true,
  allcolors=blue,
}
\usepackage{titlesec}

\titleformat{\section}
  {\normalfont\bfseries\centering}{\thesection.}{1em}{}
\begin{document}
\renewcommand{\thesection}{\arabic{section}}
\renewcommand{\thetable}{\arabic{table}}
\renewcommand{\thesubsection}{\arabic{section}.\arabic{subsection}}
\title{On the optimality of the radical-pair quantum compass} 

\author{Luke D.\ Smith\,\orcidlink{0000-0002-6255-2252}}
\affiliation{Department of Physics and Astronomy, University of Exeter, Stocker Road, Exeter EX4 4QD, UK}
\affiliation{Living Systems Institute, University of Exeter, Stocker Road, Exeter EX4 4QL, UK}

\author{Jonas Glatthard\,\orcidlink{0000-0002-8411-4958}}
\affiliation{Department of Physics and Astronomy, University of Exeter, Stocker Road, Exeter EX4 4QD, UK}

\author{Farhan T.\ Chowdhury\,\orcidlink{0000-0001-8229-2374}}
\affiliation{Department of Physics and Astronomy, University of Exeter, Stocker Road, Exeter EX4 4QD, UK}
\affiliation{Living Systems Institute, University of Exeter, Stocker Road, Exeter EX4 4QL, UK}

\author{Daniel R.\ Kattnig\,\orcidlink{0000-0003-4236-2627}}
\affiliation{Department of Physics and Astronomy, University of Exeter, Stocker Road, Exeter EX4 4QD, UK}
\affiliation{Living Systems Institute, University of Exeter, Stocker Road, Exeter EX4 4QL, UK}

\begin{abstract}
Quantum sensing enables the ultimate precision attainable in parameter estimation. Circumstantial evidence suggests that certain organisms, most notably migratory songbirds, also harness quantum-enhanced magnetic field sensing via a radical-pair-based chemical compass for the precise detection of the weak geomagnetic field. However, what underpins the acuity of such a compass operating in a noisy biological setting, at physiological temperatures, remains an open question. Here, we address the fundamental limits of inferring geomagnetic field directions from radical-pair spin dynamics. Specifically, we compare the compass precision, as derived from the directional dependence of the radical-pair recombination yield, to the ultimate precision potentially realisable by a quantum measurement on the spin system under steady-state conditions. To this end, we probe the quantum Fisher information and associated Cram\'er--Rao bound in spin models of realistic complexity, accounting for complex inter-radical interactions, a multitude of hyperfine couplings, and asymmetric recombination kinetics, as characteristic for the magnetosensory protein cryptochrome. We compare several models implicated in cryptochrome magnetoreception and unveil their optimality through the precision of measurements ostensibly accessible to nature. Overall, the comparison provides insight into processes honed by nature to realise optimality whilst constrained to operating with mere reaction yields. Generally, the inference of compass orientation from recombination yields approaches optimality in the limits of complexity, yet plateaus short of the theoretical optimal precision bounds by up to one or two orders of magnitude, thus underscoring the potential for improving on design principles inherent to natural systems.

\end{abstract}

\maketitle

\section{Introduction}
Quantum magnetometers \cite{Degen2017} offer the ability to detect weak magnetic fields with unprecedented sensitivity \cite{Giovannetti2004}. These sensors have numerous established, developing, and proposed uses, including: biomedical applications \cite{Aslam2023}, research into fundamental physics \cite{Asztalos2010}, and navigational systems \cite{Wang2023}. Remarkably, a promising hypothesis and circumstantial evidence suggests that nature may also have developed a quantum magnetometer that, relying on coherent spin dynamics in a radical-pair, permits the precise detection of the weak geomagnetic field (50 $\mu$T). Besides magnetite-based sensors \cite{Shaw2015}, this Radical-Pair Mechanism (RPM) \cite{Hore2016, Rodgers2009, Schulten1978a} is now discussed as the fundamental enabler of a magnetic field sense, magnetoreception, which is widespread in migratory animals \cite{Mouritsen2018, Lohmann2010}, such as migratory songbirds \cite{Wiltschko2019}, that use it for navigation. Unlike synthetic quantum magnetometers, nature's magnetometer is protein-based and operates at physiological temperatures (approximately 310 K in birds) in a biological environment, which presents a puzzle to its ability to deliver remarkable acuity under hot and noisy conditions.

Magnetoreception has been extensively documented through ethological observations and its quantum theoretical basis has been widely studied. The RPM supports the fundamental principles of a quantum spin-dependent chemical compass. However, many details underlying the mechanism, not least the identity of the sensory receptor \cite{Nordmann2017}, have remained elusive or are the subject of ongoing debate. Experimental observations of migratory birds under different light conditions \cite{Wiltschko2010, Wiltschko2001, Wiltschko1995} and degraded vision \cite{Stapput2010}, suggested that the mechanism might be vision-based with the receptor located in the eye. Flavin-binding cryptochrome proteins, excitable by blue light and found, for example, in avian double-cone and long-wavelength single-cone photoreceptor cells \cite{Gunther2018}, emerged as a key candidate thought to harbor the magnetosensitive radical-pair \cite{Ritz2000}. Since its original proposal, several supporting studies have emerged exploring the photochemistry of cryptochrome. For example, the cryptochrome 4 of several bird species has been established to form a photo-induced flavin adenine dinucleotide (FAD)/tryptophan radical-pair $[\mathrm{FAD}^{\bullet-}/ \mathrm{W}^{\bullet +}]$ via a sequential electron transfer reaction, which is magnetic field-sensitive \textit{in vitro} \cite{Xu2021, Hochstoeger2020}, albeit to fields exceeding the geomagnetic field by orders of magnitude.

\begin{figure}
\centering
	\includegraphics[scale=1]{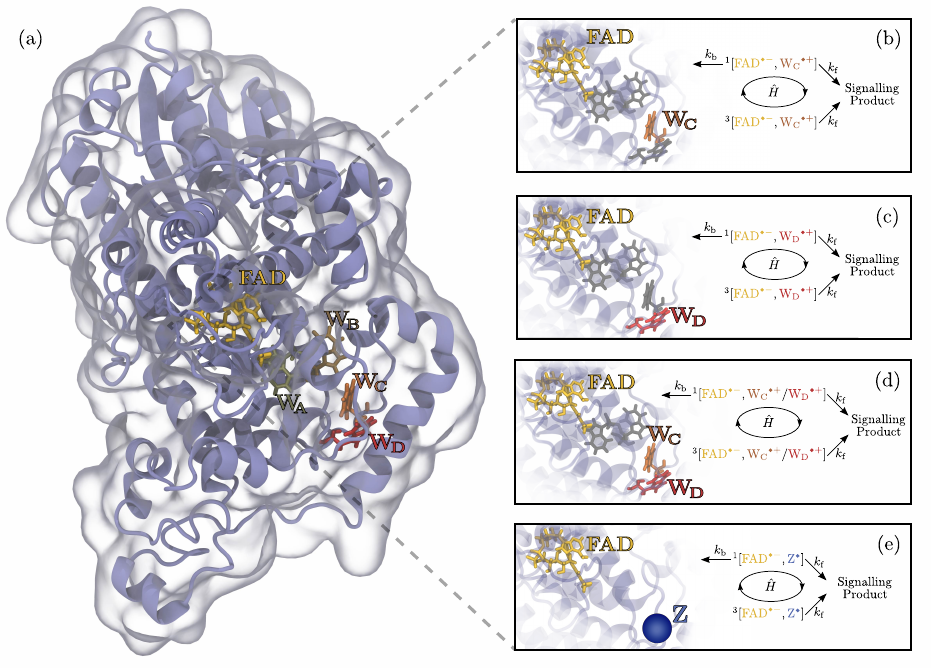}
	\caption{(a) Avian cryptochrome structure \cite{Schuhmann2021} with highlighted FAD cofactor and the four tryptophan residues labelled $\mathrm{W_{A}, W_{B}, W_{C},}$ and $\mathrm{W_{D}}$) that form the tryptophan tetrad. Photo-excitation of FAD initiates sequential electron transfer reactions along the tetrad forming a radical-pair of the form $[\mathrm{FAD}^{\bullet-}, \mathrm{W}^{\bullet +}]$. Specifically, we consider the well-separated radical-pairs (b) $[\mathrm{FAD}^{\bullet-}, \mathrm{W_{C}}^{\bullet +}]$, (c) $[\mathrm{FAD}^{\bullet-}, \mathrm{W_{D}}^{\bullet +}]$, (d) and a composite radical-pair $[\mathrm{FAD}^{\bullet-}, \mathrm{W_{C}}^{\bullet +}/\mathrm{W_{D}}^{\bullet +}]$ characterised by a fast degenerate electron transfer between $\mathrm{W_{C}}$ and $\mathrm{W_{D}}$ that averages the hyperfine interactions. For comparison, we also consider the reference-probe radical-pair $[\mathrm{FAD}^{\bullet-}, \mathrm{Z}^{\bullet}]$ (e), formed between $\mathrm{FAD}^{\bullet-},$ and a $\mathrm{Z}^{\bullet}$ radical devoid of hyperfine interactions, where the position represented by a blue sphere is arbitrarily chosen in the figure. In (b)-(e) simplified schemes of the reaction mechanisms are shown, in which the radical-pair is born in the singlet state, and can coherently interconvert via the Hamiltonian $\hat{H}$ to the triplet state, with spin multiplicity shown by superscript labels one and three, respectively. A signalling product can be formed from both singlet and triplets, with rate $k_{\mathrm{f}}$, whereas only the singlet state can recombine to form the diamagnetic resting state, with rate $k_{\mathrm{b}}$.} \label{fig.CRY_scheme}
\end{figure}

Whilst studies have demonstrated magnetic field effects \textit{in vitro} \cite{Kerpal2019, Maeda2008, Timmel1998}, the origin of magnetosensitivity has not been unequivocally determined under \textit{in vivo} conditions, and alternative radical-pairs have been proposed besides $[\mathrm{FAD}^{\bullet-}/ \mathrm{W}^{\bullet +}]$. A so-called reference-probe radical-pair \cite{Procopio2020, Cai2012, Gauger2011, Ritz2010, Timmel2001}, that boasts increased sensitivity, could be formed from FAD and a different electron donor free from hyperfine interactions $[\mathrm{FAD}^{-\bullet}/ \mathrm{Z}^{\bullet}]$\cite{Lee2014}. The dark-state reoxidation of fully reduced FAD to form a flavin semiquinone/superoxide radical-pair $[\mathrm{FADH}^{\bullet}/ \mathrm{O_{2}}^{\bullet -}]$ is frequently discussed in this context \cite{Pooam2019, Wiltschko2016, Solovyov2009, Ritz2009}. However, this radical-pair is questionable in view of the intrinsically fast spin relaxation associated with superoxide \cite{Mondal2019, Muller2011, Hogben2009}. In figure \ref{fig.CRY_scheme} we show a model of cryptochrome and several potential reaction schemes empowering the radical-pair mechanisms, as relevant to this study. More detailed schemes, explicating the chemistry associated with different radical-pair realisations, can be found in the existing literature \cite{Xu2021, Babcock2021, Hochstoeger2020, Wiltschko2016}.

Whilst the exact details of the magnetosensitive radical-pair recombination reactions in cryptochrome might differ, and indeed has not been definitively established, some essential elements are shared over several proposed mechanisms. Namely, the electron spins of the radicals interact with the geomagnetic field (or an applied magnetic field) through the Zeeman interaction, and with one another through spin-selective recombination and inter-radical interactions, such as the electron-electron dipolar (EED) and exchange interactions. In addition, at least one of the electron spins interacts with surrounding magnetic nuclei through hyperfine interactions. The Zeeman interaction together with anisotropic hyperfine interactions provide a magnetic field orientation-dependent coherent inter-conversion between singlet and triplet states of the electron spins. Note that anisotropic EED coupling alone can also achieve this \cite{Keens2018}, but only in systems of more than two radicals \cite{Babcock2021, Babcock2020,Kattnig2017, Kattnig2017a}. Magnetic field sensitivity thus arises from a difference in the branching ratio of reactants to products formed from spin-selective reaction channels, i.e.\ channels that differentiate the singlet and triplet configurations (e.g. as radical-pair recombination is only permitted in the singlet state). The EED interaction is unavoidable and limits the attainable sensitivity \cite{Babcock2020}, but has nonetheless often been neglected from the discussion.

RPM magnetoreception has likely evolved \cite{Bartolke2021, Liedvogel2010, Cashmore1999} over millions of years (birds evolved from theropod dinosaurs during the Jurassic around 165–150 million years ago), emerging in the presence of said adverse interactions, which are unavoidable and apply generally to radical mechanisms. This naturally raises fundamental questions about nature's ability to utilise quantum processes. \emph{A priori}, one would assume that, given the long-term evolutionary optimisation and the essential role of compass navigation for migratory animals, a high level of optimality should have been attained. However, do these sensors operate close to the intrinsic precision bounds of quantum processes? Conversely, could nature have developed alternative, yet unexplored, sensory processing mechanisms that provide sensitivity and precision augmentation? These are the questions that this contribution seeks to explore.

To assess the optimality of the radical-pair compass, we consider the quantum Fisher information (QFI) \cite{Braunstein1994} and associated quantum Cram\'er--Rao bound (QCRB) \cite{Holevo2011}, which constrains the maximal precision attainable in the statistical estimation of a parameter, say $\theta$, from measurements on a quantum state $\rho(\theta)$. For an unbiased estimator based on $N$ independent measurements, the bound is provided by
\begin{equation}
\frac{1}{\delta^2\theta}\leq N F(\theta)\leq N\pazocal{F}(\theta), \label{eq.cramer_rao}
\end{equation}
where $\delta^2\theta$ represents the variance of the $\theta$-estimates, $F(\theta)$ is the classical Fisher information (CFI) \cite{Fisher1925} for any specified observable, and the QFI is represented by $\pazocal{F}(\theta)$. The QFI quantifies the degree (in terms of the Uhlmann fidelity) to which the quantum state changes in response to an infinitesimal change of the parameter $\theta$. Hence, $\pazocal{F}(\theta)$ sets the precision bound of a single measurement, whereby a large QFI, i.e.\, large change of the state, corresponds to a more precise parameter estimation. The approach has analogues in classical estimation theory, the (classical) Cram\'er--Rao bound \cite{Cramer1946} and the CFI, but extends those insofar as in quantum mechanics the bound can be tightened by optimally choosing the measurement basis of the underlying measurement for a particular $\theta$ value.

By leveraging the QFI and QCRB, we address questions regarding the radical-pair quantum compass which demand insight to the fundamental limits in precision of estimating an unknown parameter, such as the orientation with respect to the magnetic field. Information theoretic approaches have proved fruitful in investigating features of radical-pairs such as the fundamental limits of sensitivity under dim light relevant for nocturnal migratory species \cite{Hiscock2019}, as well as providing insight into other quantum phenomena such as quantum coherence \cite{Kominis2023, Poonia2023, Jain2021, Kominis2020, Le2020, Cai2013, Hogben2012, Gauger2011}, where we have previously demonstrated the importance of treatments comprising many nuclear spins inherent to biological systems \cite{Smith2022, Atkins2019}. The QFI has previously been utilised to interrogate the fundamental quantum limits of magnetic field sensitivity of radical-pairs in model studies of sensing the magnitude \cite{Vitalis2017} and direction of the field \cite{Guo2017}. Whilst these studies have provided foundational insight into the precision capabilities of a radical-pair compass, they are restricted to a toy model comprised of a single nuclear spin and neglected inter-radical interactions and asymmetric reaction kinetics. As such, by avoiding the complexities inherent in a biological context, i.e.\ a large number of hyperfine interactions and the presence of inter-radical interactions, these prior studies are unable to assess emergent features and answer the question of how optimal is a radical mechanism based compass of nature. Our study aims to address this missing piece by extending the system to mirror nature's complexity and adverse interactions.

\section{Methods} \label{sec.Methods}
To accommodate for the complexity and variability inherent in nature, we consider radical-pair models emulating cryptochrome 4a from the European robin (\textit{Erithacus rubecula}; ErCry4a) and cryptochrome 1 from the thale cress (\textit{Arabidopsis thaliana}; AtCry1). Our models incorporate up to 10 nuclear spins; the hyperfine interactions of which have been taken from \cite{Gesa2022} and are averaged over the thermally excited dynamical degrees of freedom of the proteins. We treat the idealised and predominantly considered scenario of neglecting EED interactions alongside the realistic, but still scarcely studied, case of taking them into account, and investigate the limit of an increasing number of hyperfine interactions. For the bird cryptochrome ErCry4a, which possesses a tryptophan tetrad (Fig.\ \ref{fig.CRY_scheme}), we investigate three FAD/tryptophan radical-pair models $[\mathrm{FAD}^{\bullet-}, \mathrm{W}^{\bullet +}]$, namely,  $[\mathrm{FAD}^{\bullet-}, \mathrm{W_{C}}^{\bullet +}]$ (denoted ErC), $[\mathrm{FAD}^{\bullet-}, \mathrm{W_{D}}^{\bullet +}]$ (denoted ErD), with $\mathrm{W_{C}}$ and $\mathrm{W_{D}}$ denoting the third and fourth tryptophan of the tryptophan tetrad, and a model referred to as $[\mathrm{FAD}^{\bullet-}, \mathrm{W_{C}}^{\bullet +}/\mathrm{W_{D}}^{\bullet +}]$ (ErC/ErD in short), for which we assume a fast degenerate electron transfer between $\mathrm{W_{C}}$ and $\mathrm{W_{D}}$ that averages the hyperfine interactions, as suggested in \cite{Gesa2022,Xu2021,Wong2021a}. For AtCry1, which lacks $\mathrm{W_{D}}$, we investigate $[\mathrm{FAD}^{\bullet-}, \mathrm{W_{C}}^{\bullet +}]$ under the label AtC. For comparison, we also investigate the reference-probe type radical-pair $[\mathrm{FAD}^{\bullet-}, \mathrm{Z}^{\bullet}]$, for which the flavin radical forms a pair with a radical, $\mathrm{Z}^{\bullet}$, devoid of hyperfine interactions. The different models are shown in figure \ref{fig.CRY_scheme}.  

By calculating the QFI of the reduced electronic state of these models under steady excitation, we can unveil their optimality in terms of their precision. This allows us to analyse how optimal nature's recombination-yield-based measurements are, assess models relative to one another, and identify how much room for improvement exists. Pertinent features of the theoretical model and the simulation approach are presented below.
 
\subsection{Radical-pair hamiltonian}
We consider radical-pairs subject to Zeeman, hyperfine, and electron-electron dipolar interactions (with $\hbar  = 1$), for which the Hamiltonian takes the form
\begin{align}
    \hat H = {\hat H_{\rm{A}}} + {\hat H_{\rm{B}}} + \hat{H}_{\mathrm{dip}}, \label{eq.Hamiltonian}
\end{align}
where the Zeeman and hyperfine interactions are accounted for by $H_{i}$ for $i \in \{\mathrm{A, B}\}$, where subscripts A and B correspond to each radical in the pair, and take the form
\begin{align}
    {\hat H_{\rm{i}}} = \sum\limits_j {{{{\bf{\hat S}}}_i} \cdot {{\bf{A}}_{i,j}} \cdot {{{\bf{\hat I}}}_{i,j}}}  + {\mathbf \omega _i} \cdot {{\bf{\hat S}}_i}.
\end{align}
In the above $\hat{\mathbf{S}}_{i}$ represents the electron spins and couple via the Zeeman interaction to the magnetic field $\mathbf{B} = B_{0}(\sin \theta \cos \phi, \sin \theta \sin \phi, \cos \theta)$ with Larmor precession angular frequency $\vec{\omega}_{i} = - \gamma_{i}\mathbf{B}$, for which $\gamma_{i} =\mu_{B}g_i$ denotes the gyromagnetic ratio of the electron in radical $i$, $\mu_{B}$ the Bohr magneton, and $g_i\approx 2.0013$ the electron $g$-factor. The nuclear spins are denoted $\hat{\mathbf{I}}_{i,j}$ and couple to the electron spins via the hyperfine interaction with coupling characterised by the hyperfine coupling tensor ${{\bf{A}}_{i,j}} $. By explicitly treating both nuclear and electron spins as part of the system, we account for the nuclear-electron spin correlations, which have been shown to be relevant to magnetic field effects \cite{Smith2022, Cai2013}. We thus explicitly model effects that would appear as non-Markovian system-bath coupling if the nuclear spins were treated as an environment. 
The EED interactions are given by the Hamiltonian $\hat{H} = {{{\bf{\hat S}}}_\mathrm{A}} \cdot {\bf D} \cdot {{{\bf{\hat S}}}_\mathrm{B}}$
where ${\bf D}$ is the EED tensor.
Specific details of the hyperfine and EED tensors used in calculations, can be found in tables S1-S6 of the Supplementary Materials (SM). 

\subsection{Spin dynamics and calculation of probe state}
We assume that the radical-pair is formed initially in a singlet state, given by a spin density operator $\hat{\rho}(0) = \frac{\hat{P}_{\mathrm{S}}}{Z},$ where $\hat{P}_{\mathrm{S}}$ is the singlet projection operator and $Z=Z_{\mathrm{A}}Z_{\mathrm{B}}$ denotes dimension of the nuclear subspace associated with the two radicals. This initial state evolves under the Hamiltonian described in eq.\ \ref{eq.Hamiltonian} via the master equation
\begin{align}
    \frac{\mathrm{d}{\hat{\rho}_{\theta}}(t)}{\mathrm{d}t} = -i[\hat{H}, \hat{\rho}_{\theta}(t)] - \frac{k_{\mathrm{b}}}{2}\{\hat{P}_{\mathrm{S}}, \hat{\rho}_{\theta}(t)\} - k_{\mathrm{f}}\hat{\rho}_{\theta}(t), \label{eq.master_equation}
\end{align}
where $k_{\mathrm{b}}$ and $k_{\mathrm{f}}$ represent the reaction rates for recombination and product formation (see figure \ref{fig.CRY_scheme}), the commutator and anticommutator are denoted by $[\cdot, \cdot]$ and $\{\cdot, \cdot\}$, respectively, and $\hat{\rho}_{\theta} = \hat{\rho}(t, \theta, \phi)$ denotes that the state acquires sensitivity to the magnetic field parameters through the Zeeman interaction. Throughout this work we use $k_{\mathrm{b}}=k_{\mathrm{f}}=1\mu$s$^{-1}$, reflecting the asymmetry in the reaction channels from singlet and triplet states typically found in cryptochromes.  

A representative probe state is necessary to evaluate the precision achievable in parameter estimation by a radical-pair system. We choose the radical-pair in a steady-state of continuous generation and recombination, as it closely resembles the natural conditions of magnetoreception. Based on this probe state we assess how precisely the deflection of the magnetic field direction from the normal of the FAD ring plane, identified with $\theta$, may be estimated. With respect to this choice, the dominant hyperfine interactions vary maximally as $\theta$ is changed. To obtain the probe state, we first introduce the Green's superoperator as the solution of 
\begin{align}
    \frac{\mathrm{d}}{\mathrm{d}t}  \hat{\hat{G}}(t, t_{0}) + i \hat{\hat{L}}(t, t_{0})\hat{\hat{G}}(t,t_{0}) = \delta(t-t_{0}), 
\end{align}
where $\hat{\hat{L}}$ is the Liouvillian accounting for coherent evolution under the Hamiltonian $\hat{H}(t,t_{0})$ and spin-selective recombination, expressed through $\hat{K}(t, t_{0})$, i.e.\ $\hat{\hat{L}}[\hat{\rho}_{\theta}] = [\hat{H},\hat{\rho}_{\theta}] - i \{ \hat{K}, \hat{\rho}_{\theta} \} $. If radical-pairs are generated at a rate and in a state described by $\hat{R}(t)$, the concentration-weighted spin density operator of the radical-pair at time $t$ is obtained from
\begin{align}
    \hat{\rho}_{\theta}(t) = \int_{-\infty}^{t} \hat{\hat{G}}(t, t_{0})\hat{R}(t_{0}) \, \mathrm{d}t_{0}.
\end{align}
Formally, $\hat{\hat{G}}$ can be expressed in terms of an exponential operator as 
\begin{align}
    \hat{\hat{G}}(t,t_{0}) = \hat{T}\exp{\Bigg(-i \int_{t_{0}}^{t} \hat{\hat{L}}(\tau, t_{0}) \, \mathrm{d}\tau \Bigg)},
\end{align}
where $\hat{T}$ is the Dyson time-ordering operator. By further assuming that the Hamiltonian and recombination are dependent on the time since radical generation, but are independent of absolute time, i.e.\ $\hat{H}(t, t_{0}) = \hat{H}(t-t_{0})$, we obtain $\hat{\hat{G}}(t, t_{0}) = \hat{\hat{G}}(t-t_{0})$.

Here, we assume that radical-pairs are generated at a constant rate in the singlet state, i.e.\ $\hat{R}(t) = \hat{R}_{0} =  k_{0} c \hat{P}_{\mathrm{S}}/Z$, with $k_{0}$ denoting the generation rate and $c$ the (constant) concentration of cryptochrome in the resting state. This assumption is valid at low excitation rates, for which the cryptochrome resting-state population is not significantly depleted by excitation. If the excitation rate is increased, the initial state could be altered through accumulation of nuclear polarisation as described in \cite{Wong2021}. For $\hat{R}(t) = \hat{R}_{0}$, $\hat{\rho}_{\theta}(t)$ assumes a steady-state given as the integral of states at different evolution times, $\hat{\hat{G}}(t-t_{0})\hat{R}_{0}$, explicitly 
\begin{align}
    \hat{\rho}_{\mathrm{ss}} &= \int_{-\infty}^{t} \hat{\hat{G}}(t-t_{0}) \hat{R}_{0} \, \mathrm{d}t_{0} \nonumber\\ &= \frac{k_{0}c}{Z} \int_{-\infty}^{t}\hat{\hat{G}}(t-t_{0}) \hat{P}_{\mathrm{S}} \, \mathrm{d}t_{0} \nonumber \\ &= \frac{k_{0}c}{Z}\int_{0}^{\infty} \hat{\hat{G}}(t)\hat{P}_{\mathrm{S}} \, \mathrm{d}t \nonumber \\  &= \frac{k_{0}c}{Z} \Tilde{G}(s=0)\hat{P}_{\mathrm{S}},
\end{align}
where $\Tilde{G}(s)$ is the Laplace transform of $\hat{\hat{G}}(t)$, which for time-independent Hamiltonians is given by $\Tilde{G}(s) = (\hat{\hat{L}} + \pmb{1}s)^{-1}$. We postulate that the electronic part of the steady-state, which has imprinted the information on the direction and intensity of the magnetic field, and which is partly inferable through spin-chemical reactions, represents the quantum spin state of cryptochrome relevant to magnetoreception. This view is in line with the approach in \cite{Guo2017}, to which it reduces to in the case of homogeneous recombination, i.e.\ the idealised picture of recombination and product formation proceeding at the same rate from the singlet and triplet state. With $\hat{\rho}_{\theta}(t)$ is associated the recombination flux $F_{\mathrm{r}} = k_{\mathrm{r}} \mathrm{Tr}[\hat{P}_{\mathrm{S}} \hat{\rho}_{\mathrm{ss}}]$. Traditionally, the change of this flux with the direction of the magnetic field is considered as decisive in determining the compass sensitivity, whereby the maximal change of the singlet recombination quantum yield, $\Phi_{\mathrm{S}} = F_{\mathrm{r}}/k_{0}c$, is a widely applied fidelity measure.
Though some works have also considered the case of access to time-resolved measurements of sensitivity \cite{Vitalis2017}, we choose to employ the steady-state picture as it more closely corresponds to the physiological conditions of magnetoreception under constant light excitation and allows a formal rationalisation of the form of the steady-state, to which the quantum parameter estimation approaches are applied. 

By calculating the average singlet yield over all directions of the magnetic field $\mathbf{B}(\phi, \theta)$ via
\begin{align}
    \overline{\Phi}_{\mathrm{S}} = \frac{1}{2\pi}\int_{0}^{\pi} \int_{0}^{\pi} \sin(\theta)\Phi_{\mathrm{S}} \mathbf{B} \, \mathrm{d\theta \, d\phi},
\end{align}
along with the maximum and minimum singlet yields $\Phi_{\mathrm{S}, \mathrm{max}}$ and $\Phi_{\mathrm{S}, \mathrm{min}}$, the anisotropy, a commonly implemented measure of sensitivity that quantifies the contrast in reaction yields over all magnetic field directions, is given by
\begin{align}
    \Gamma = (\Phi_{\mathrm{S}, \mathrm{max}} - \Phi_{\mathrm{S}, \mathrm{min}})/\overline{\Phi}_{S} \label{eq.anisotropy}.
\end{align}
The anisotropy thus quantifies the sensitivity by assessing whether a contrast is produced in the branching ratio of singlets to triplets, a small value suggesting there is a small response to the magnetic field; a large value indicates a marked change in the branching ratio when rotating the magnetic field between the two extremal orientations. 

\subsection{Quantum Fisher information and optimal measurements}
In quantum metrology, a quantum system encodes a parameter $\theta$ in its state $\hat{\rho}_\theta$. The parameter $\theta$ is to be estimated from measurements on the system. The statistical uncertainty of an unbiased estimator is bounded by the quantum Fisher information $\pazocal{F}_\theta$ through the Cram\'er--Rao inequality (eq.\ \ref{eq.cramer_rao}). 
Specifically, the QFI is given by \cite{Braunstein1994}
\begin{equation}
	\pazocal{F}_\theta = -2 \lim_{ \tau \rightarrow 0} \frac{\partial^2  \mathbb{F} (\hat{\rho}_{\theta-\tau/2},\hat{\rho}_{\theta+\tau/2})}{\partial \tau^2}, \label{eq.QFI}
\end{equation}
where the Uhlmann fidelity $ \mathbb{F}(\hat{\rho}_1,\hat{\rho}_2) $, defined as
\begin{equation}
	\mathbb{F} (\hat{\rho}_1,\hat{\rho}_2) = \mathrm{Tr}\left(\sqrt{\sqrt{\hat{\rho}_1}\,\hat{\rho}_2\, \sqrt{\hat{\rho}_1}}\right)^2,
\end{equation}
provides a measure of the closeness of the two states $\hat{\rho}_1,\hat{\rho}_2$.
Thus, it bounds the precision, with which the parameter $\theta$ can be estimated from measurement on the system and encodes the sensitivity of the state to the parameter. The overall sensitivity of the state can be compared to the sensitivity obtained with a specific measurement. This is given by the classical Fisher information \cite{Fisher1925,Paris2011}, which here takes the form
\begin{align}\label{eq.defcfi}
F_\theta &= \sum\nolimits_n p_{n,\theta}\, (\partial_\theta \ln p_{n,\theta})^2 \nonumber\\
&= \sum_{n} \frac{(\partial_{\theta}p_{n,\theta})^{2}}{p_{n,\theta}},
\end{align}
where we have used $\partial_{\theta}$ as shorthand for $\mathrm{d}/\mathrm{d\theta}$, and $p_{n,\theta} = \mathrm{Tr}[\hat{\rho}_\theta \hat{\Pi}_{n}]$ are the Born rule probabilities associated with the measurement element $\hat{\Pi}_{n}$, where $\sum_{n} \hat{\Pi}_{n} = \pmb{1}$, given that the state of the system before the measurement was $\hat{\rho}_\theta$. The QFI is obtained from optimising the CFI over all possible measurements.

Thus, noting that $\partial_{\theta}p_{n,\theta} = \mathrm{Tr}[\partial_{\theta} \hat{\rho}_\theta \hat{\Pi}_{n}] = \mathrm{Tr}[\hat{\rho}_\theta \hat{\Pi}_{n}\hat{L}_{\theta}]$, and maximising the CFI over $\{ \hat{\Pi}_{n} \}$, the QFI is obtained as \cite{Paris2011}
\begin{align}
    \mathrm{Tr} \left( \hat{L}_{\theta}^2 \hat{\rho}_\theta \right)= \pazocal{F}_\theta, \label{eq.qfi_sld}
\end{align}
where $\hat{L}_{\theta}$ is the symmetric logarithmic derivative, which is implicitly defined by the Lyapunov equation \cite{Helstrom1969, Holevo2011, Braunstein1994}
\begin{equation}\label{eq.defsld}
    \partial_\theta\, \hat{\rho}_\theta = \frac{1}{2}\lbrace \hat{L}_{\theta}, \hat{\rho}_\theta \rbrace,
\end{equation}
where $\{\cdot, \cdot\}$ denotes the anticommutator. This equation is uniquely solved by \cite{Bhatia1997}
\begin{align}\label{eq.lyapunov}
\hat{L}_{\theta} = 2\int^\infty_0 e^{-\lambda \hat{\rho}_\theta} (\partial_\theta \, \hat{\rho}_\theta) e^{-\lambda \hat{\rho}_\theta} \, d\lambda,
\end{align}
which satisfies $\mathrm{Tr} \left( \hat{L}_{\theta} \hat{\rho}_\theta \right)= 0$. 
In the basis that diagonalises $\hat{\rho}_{\theta}$, the QFI can be equivalently evaluated via eq.\ \ref{eq.qfi_sld} by utilising the spectral decomposition ${\hat{\rho}_\theta = \sum_{i} p_{i} \vert \psi_{i} \rangle \langle \psi_{i} \vert}$ to obtain \cite{Braunstein1994}
\begin{align}
    \pazocal{F}_\theta = 2 \sum_{p_{i}+p_{j} \neq 0} \frac{1}{p_{i} + p_{j}} \big\vert \langle \psi_{i} \vert \partial_{\theta}\hat{\rho}_{\theta} \vert \psi_{j} \rangle \big \vert^{2}.
\end{align}
If $\hat{\rho}_{\theta}$ is invertible, $\pazocal{F}_\theta$ given by \cite{Safranek2018} can furthermore be obtained from
\begin{align}\label{eq.sqfi}
\pazocal{F_{\theta}} = \mathrm{vec}\left( \partial_\theta \hat{\rho}_\theta \right)^\dagger \mathrm{vec}(\hat{L}_{\theta}),
\end{align}
where $\mathrm{vec}(\cdot)$ denotes vectorisation of a matrix and 
\begin{align}\label{eq.ssld}
\mathrm{vec}\left(  \hat{L}_{\theta} \right) = 2 \left( \overline{ \hat{\rho}}_\theta \otimes  \pmb 1 +  \pmb 1 \otimes \hat{\rho}_\theta \right)^{-1} \mathrm{vec}\left( \partial_\theta \hat{\rho}_\theta \right), 
\end{align}
where $\cdot^\dagger$ and $\overline \cdot$ stand for Hermitian transpose and complex conjugation.

The symmetric logarithmic derivative determines the optimal measurement basis, but leaves open the question of constructing an estimator which saturates the Cram\'er--Rao bound. This can be constructed by measuring the observable \cite{Paris2011, Sidhu2019}
\begin{equation}\label{eq.opt_est}
    \hat{M}_{\theta} = \theta \, \pmb 1 + \hat{L}_{\theta}/\pazocal{F}_\theta.
\end{equation}
The measurement average is $\theta$ and the variance is given by $\pazocal{F}_\theta^{-1}$. The observable depends on the variable which is to be estimated, but can be constructed adaptively \cite{Paris2011}.

\subsection{Calculating the precision of yield-based measurements}
Whilst $\pazocal{F}_\theta$ provides an upper bound of precision, one which is saturated by the optimal estimator, the consensus is that the radical-pair acquires sensitivity to magnetic fields due to a change in reaction yields (as accounted for by the anisotropy in equation \ref{eq.anisotropy}). In fact, it is unlikely that the optimal estimator is attainable within the constraints posed by biology and radical chemistry. As such, we assess precision in the estimation of a magnetic field parameter $\theta$ as realisable in nature by considering the singlet recombination yield. A recombination event corresponds to a Bernoulli trial leading either to the singlet or triplet reaction product with probability $\Phi_{\mathrm{S}}$ and $(1-\Phi_{\mathrm{S}})$, respectively \cite{Hiscock2019}. For $N$ trials, a binomial distribution thus results. The associated variance is given by $N\Phi_{\mathrm{S}}(1-\Phi_{\mathrm{S}})$, and the error in estimating $\theta$ from the mean yield is obtained by propagation of variances to give
\begin{align}
    \Delta^{2}\theta = \frac{\Phi_{\mathrm{S}}(1-\Phi_{\mathrm{S}})}{N\vert\mathrm{\frac{d\Phi_{\mathrm{S}}}{\mathrm{d}\theta}\vert^{2}}} \label{eq.variance}.
\end{align}
In Appendix \ref{sec.appendix_1} we show that the same expression can be obtained as the error associated with measuring the square of the total angular momentum $\hat{S}^{2}$ on the probe state. Specifically, the radical-pair recombination can be viewed as a measurement of the total electronic spin \cite{Guo2017}. In Appendix \ref{sec.appendix_2} we further demonstrate that $\Delta^{2}\theta$ saturates the Cram\'er-Rao bound set by $F_{\theta}$ with measurement elements $\Pi_{n} \in \{ \hat{P}_{\mathrm{S}}, \hat{P}_{\mathrm{T}} \}$, where $\hat{P}_{\mathrm{S}}$ and  $\hat{P}_{\mathrm{T}}=\pmb{1}-\hat{P}_{\mathrm{S}}$ are the singlet and triplet projection operators, respectively. Note that \cite{Vitalis2017} has suggested a Poisson binomial distribution rather than a binomial distribution from an analysis of the time-dependence of the reaction yield. However, this result appears to imply that time-dependent recombination statistics are accessible, which we do not assume is possible here.

\section{Results}
We consider the contrast in yields, and precision in estimating parameter $\theta$ of the magnetic field for the four radical-pair systems ErC, ErD, ErC/ErD and AtC introduced in section \ref{sec.Methods} for a $[\mathrm{FAD}^{\bullet-}, \mathrm{W}^{\bullet +}]$ model where both radicals are coupled to nuclei through hyperfine interactions, and the reference-probe model $[\mathrm{FAD}^{\bullet-}, \mathrm{Z}^{\bullet}]$, where only the FAD radical is hyperfine-coupled. All cases are considered both with and without the inclusion of EED interactions; the former is a popular abstraction, while the latter case is representative of the true system. To understand trends in the limits of biological complexity, for which many nuclear hyperfine-interactions are present, we explore the effect of incrementally adding hyperfine-interactions in order of decreasing importance, as assessed by the hyperfine tensors largest magnitude eigenvalue (see figures S1-S4 of the SM). In all simulations, we sample across orientation parameters of the magnetic field at a resolution of 1 degree in $\theta$, which has associated with it the most significant changes in the hyperfine field of the dominant hyperfine interactions, in particular N5 and N10, and up to 5 degrees in $\phi$. Further computational details are provided in the SM.

\begin{figure}[t!]
\centering
	\includegraphics[scale=1]{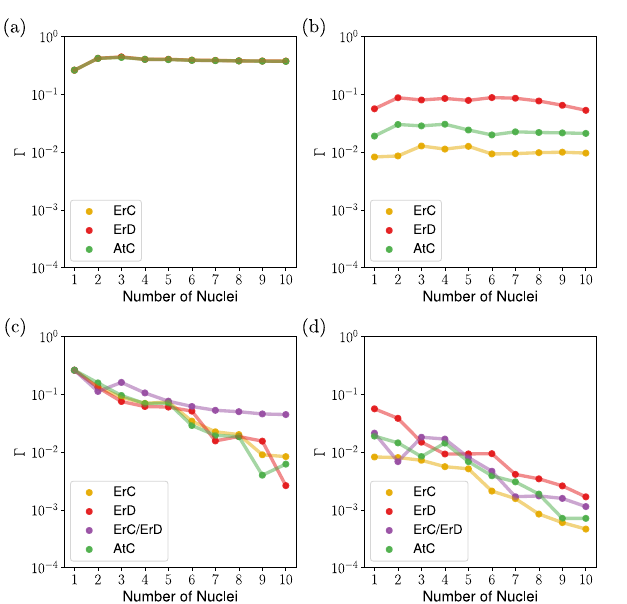}
	\caption{Singlet yield anisotropies $\Gamma$ for radical-pairs as the number of hyperfine coupled nuclei is increased up to ten. For bird cryptochrome (\textit{Erithacus rubecula} ErCry4a), we consider three radical-pair models ErC, ErD and ErC/ErD, involving tryptophans $\mathrm{W_{C}}$ or $\mathrm{W_{D}}$ or both, respectively. We also consider the plant cryptochrome AtCry1 in the form of AtC. Reference-probe $[\mathrm{FAD}^{\bullet-}, \mathrm{Z}^{\bullet}]$ radical-pairs are considered without EED in (a) and including EED interactions in (b). FAD/tryptophan $[\mathrm{FAD}^{\bullet-}, \mathrm{W}^{\bullet +}]$ radical-pairs are considered without EED in (c) and including EED interactions in (d).}
        \label{fig.anisotropy}
\end{figure}

In figure \ref{fig.anisotropy} we assess the effect of increasing the number of hyperfine interactions on the sensitivity measure $\Gamma$, defined in equation \ref{eq.anisotropy}, which represents the contrast in the singlet yields over all orientations to the magnetic field. For reference-probe type systems we see a small gain in the anisotropy as the number of hyperfine interactions is increased to 2-3 nuclei before it approximately plateaus. In the absence of EED interaction, the differences between ErC, ErD and AtC are minor, whereas in the case of their inclusion, ErD is observed to have the largest anisotropy. This is expected as it possesses the weakest EED interaction on account of the largest separation of radicals among the three cases. Overall, in line with previous investigations of the no-EED case \cite{Procopio2020, Kattnig2016a}, it is seen that the anisotropy is robust to the inclusion of more hyperfine-interactions. This holds true here, even in the presence of EED interactions, albeit with an overall decrease of the anisotropy.

For $[\mathrm{FAD}^{\bullet-}, \mathrm{W}^{\bullet +}]$ radical-pairs, increasing the number of nuclei generally leads to a fall in the anisotropy, in line with previous observations for ErC, ErD and AtC in the absence of EED \cite{Gesa2022}. It is noteworthy, however, that without EED interactions the composite system ErC/ErD has near an order of magnitude larger anisotropy and exhibits a trend suggesting a plateauing of the anisotropy as the number of nuclei is increased. With the inclusion of EED interactions, we observe an overall decrease in the anisotropy values, where ErD is the least affected. Notably, in the case of ErD with 1 hyperfine interaction the inclusion of EED significantly reduces the anisotropy, but as the number of nuclei is increased to 10, the anisotropy with or without EED interactions becomes comparable. The composite system falls below ErD when EED interactions are included, yet still benefits from an in part reduction in the EED interaction received from the involvement of the distant $\mathrm{W_{D}}$ tryptophan. Additionally, the property of plateauing is partly maintained across the increase from 7-10 nuclei, and as such the anisotropy approaches that achieved by ErD. 

\begin{figure}[t!]
\centering
	\includegraphics[scale=1]{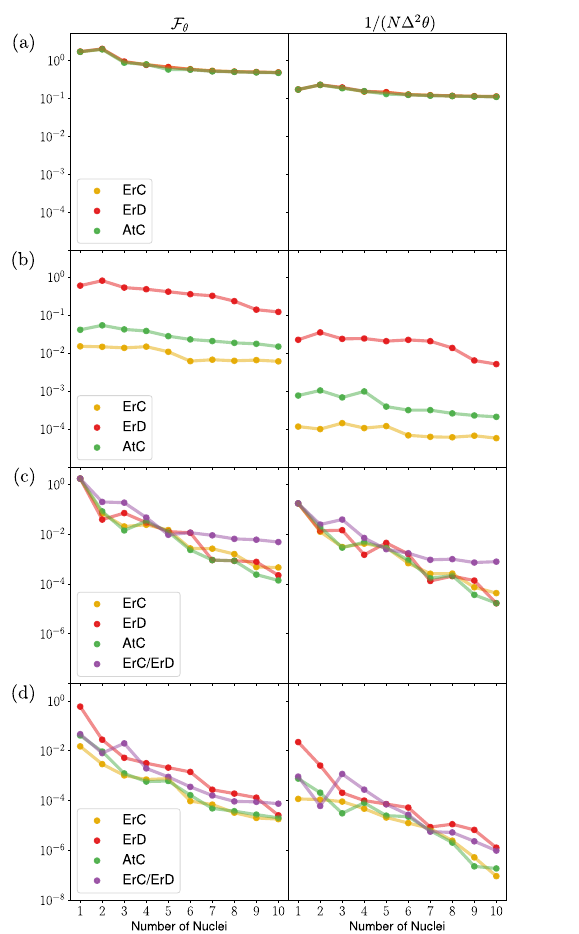}
	\caption{Quantum Fisher information $\pazocal{F}_{\theta}$, and variance in magnetic field parameter $\theta$ arising from a yield-based measurement $1/(N\Delta^{2}\theta)$, are shown at the value of $\theta$ that minimises the variance and for the best $\phi$ axis. Here, we use equivalent models to figure \ref{fig.anisotropy}. Reference-probe $[\mathrm{FAD}^{\bullet-}, \mathrm{Z}^{\bullet}]$ radical-pairs are considered without EED in (a) and including EED interactions in (b). FAD/tryptophan $[\mathrm{FAD}^{\bullet-}, \mathrm{W}^{\bullet +}]$ radical-pairs are considered without EED in (c) and including EED interactions in (d). Optimal precision in estimating $\theta$ is achieved in the limit of $N\pazocal{F}_{\theta}\cdot \Delta^{2}\theta$ approaching one, where pertinent values can be found in table \ref{table.optimality}.}
    \label{fig.precision_hyperfine}
\end{figure}

\begin{figure}[t!]
\centering
	\includegraphics[scale=1]{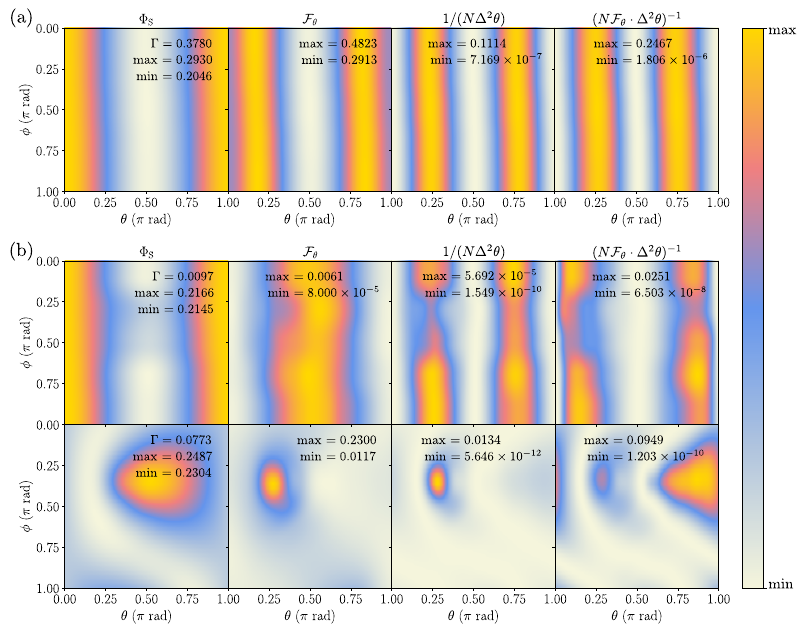}
	\caption{Singlet yield $\Phi_{\mathrm{S}}$, quantum Fisher information $\pazocal{F}_{\theta}$, variance in magnetic field parameter $\theta$ arising from a yield-based measurement $1/(N\Delta^{2}\theta)$ , and the product $(N\pazocal{F}_{\theta} \cdot \Delta^{2}\theta)^{-1}$, are shown over $\theta$ and $\phi$ parameters of the magnetic field for select examples of $[\mathrm{FAD}^{\bullet-}, \mathrm{Z}^{\bullet}]$ radical-pairs. (a) ErC with $N_{\hat{I}}=10$ nuclei neglecting EED. (b) ErC with $N_{\hat{I}}=10$ (top), and ErD with $N_{\hat{I}}=8$ (bottom) nuclei including EED interactions. Maximum and minimum attained values, and the singlet yield anisotropy $\Gamma$, are also presented.}
    \label{fig.heatmaps_hyperfine_1}
\end{figure}

\begin{figure}[t!]
\centering
	\includegraphics[scale=1]{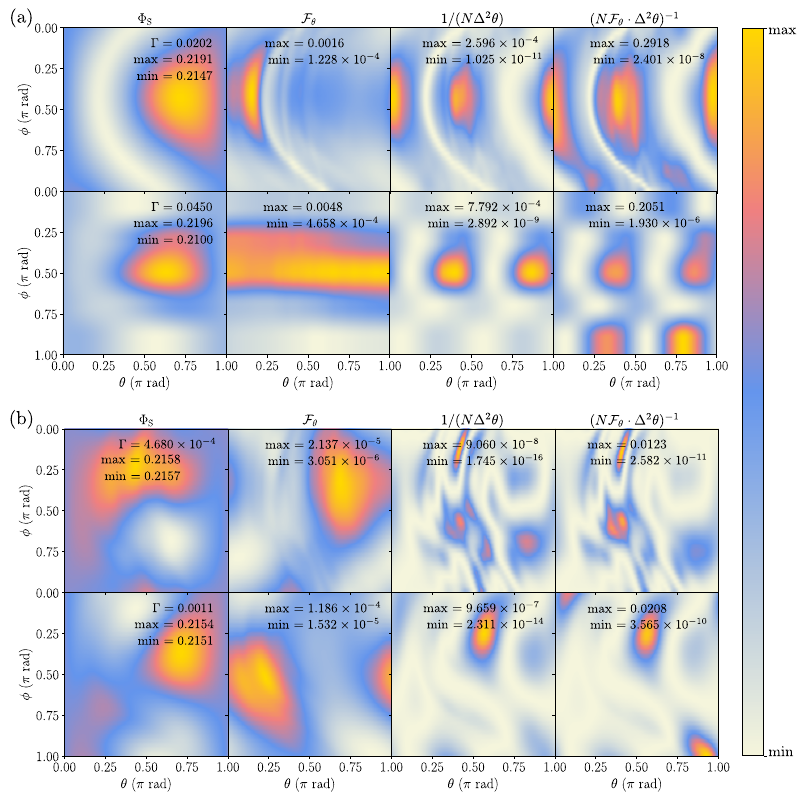}
	\caption{Singlet yield $\Phi_{\mathrm{S}}$, quantum Fisher information $\pazocal{F}_{\theta}$, variance in magnetic field parameter $\theta$ arising from a yield-based measurement $1/(N\Delta^{2}\theta)$ , and the product $(N\pazocal{F}_{\theta} \cdot \Delta^{2}\theta)^{-1}$, are shown over $\theta$ and $\phi$ parameters of the magnetic field for select examples of $[\mathrm{FAD}^{\bullet-}, \mathrm{W}^{\bullet +}]$ radical-pairs. (a) ErC with $N_{\hat{I}}=8$ nuclei (top) and ErC/ErD with $N_{\hat{I}}$ nuclei (bottom) neglecting EED. (b) ErC with $N_{\hat{I}}=10$ nuclei (top), and ErC/ErD with $N_{\hat{I}}=10$ nuclei (bottom) including EED interactions. Maximum and minimum attained values, and the singlet yield anisotropy $\Gamma$, are also presented.}
    \label{fig.heatmaps_hyperfine_2}
\end{figure}

In figure \ref{fig.precision_hyperfine} we assess the precision of the radical-pair systems via the quantum Fisher information $\pazocal{F}_{\theta}$, which provides a bound (QCRB) of $\Delta^{2}\theta$ in the estimation of the magnetic field direction as expressed through $\theta$ (cf.\ eq.\ \ref{eq.cramer_rao}). We analyze how closely the variance of $\theta$ as resulting from yield measurements, eq.\ \ref{eq.variance}, approaches the QCRB. We choose to assess these quantities here in terms of the maximum precision attainable in estimating $\theta$ as a function of $\theta$ and $\phi$, as assessed by $1/(N\Delta^{2}\theta)$, and comparing it to the value of $\pazocal{F}_{\theta}$ for the same magnetic field orientation. In the SM (figures S5-S7), we provide a range of alternative assessments, such as taking the average over $\theta$, and comparing the maximum achieved independently by $\pazocal{F}_{\theta}$ and $1/(N\Delta^{2}\theta)$, where similar observations emerge. To assess how optimal the precision of a yield-based measure is, we evaluate $N\pazocal{F}_{\theta} \cdot \Delta^{2}\theta$. Table \ref{table.optimality} summarises the highest and lowest optimality attained, the average over $N_{\hat{I}}$, and for $N_{\hat{I}}=1$, where $N_{\hat{I}}$ is the number of considered hyperfine-coupled  nuclei. A value of $N\pazocal{F}_{\theta} \cdot \Delta^{2}\theta$ of 1 would saturate the QCRB (cf.\ eq.\ \ref{eq.cramer_rao}) and identify that the measure is optimally precise. In figure \ref{fig.heatmaps_hyperfine_1} and figure \ref{fig.heatmaps_hyperfine_2} we present how $\Phi_{\mathrm{S}}$, $\pazocal{F}_{\theta}$, $1/(N\Delta^{2}\theta)$ and $N\pazocal{F}_{\theta} \cdot\Delta^{2}\theta$ vary over $\theta$ and $\phi$, for $[\mathrm{FAD}^{\bullet-}, \mathrm{Z}^{\bullet}]$ and $[\mathrm{FAD}^{\bullet-}, \mathrm{W}^{\bullet +}]$ systems, respectively, to elucidate finer details for select examples. The complete set of heatmaps for $N_{\hat{I}}=1$--$10$ can be found in the SM (figures S8--S21). 

In the case of $[\mathrm{FAD}^{\bullet-}, \mathrm{Z}^{\bullet}]$ without EED interactions, shown in figure \ref{fig.precision_hyperfine}(a), we see similar trends emerge as was found for the anisotropy of the yield. Namely, there is little difference between ErC, ErD and AtC, and the best $1/(N\Delta^{2}\theta)$ and associated $\pazocal{F}_{\theta}$ initially increase as the number of nuclei is incremented from 1-2, but gradually decline thereafter. A rise is seen in ${N\pazocal{F}_{\theta} \cdot \Delta^{2}\theta}$ as $N_{\hat{I}}$ is incremented that plateaus at approximately 4. For the characteristic parameters as a function of $\theta$ and $\phi$, we observe that ErC, ErD, and AtC share similar behaviour and that little changes as the number of nuclei is increased. In figure \ref{fig.heatmaps_hyperfine_1}(a) we show a representative example, namely, ErC with $N_{\hat{I}}=10$ nuclei, which demonstrates that any choice of $\phi$ provides a scanning axis with near to the maximal contrast in $\Phi_{\mathrm{S}}$ attained.

\begin{table}
\centering
\caption{$\pazocal{F}_{\theta} \cdot \Delta^{2}\theta$ is reported rounded to integer value for: $N_{\hat{I}}=1$, the lowest optimality attained as represented by the max value, the highest optimality attained as represented by the min value, and a robust average formed from the 7 $N_{\hat{I}}$ with smallest $\pazocal{F}_{\theta} \cdot \Delta^{2}\theta$. Reference-probe $[\mathrm{FAD}^{\bullet-}, \mathrm{Z}^{\bullet}]$ and $[\mathrm{FAD}^{\bullet-}, \mathrm{W}^{\bullet +}]$ radical-pairs are compared against one another, for the case of neglecting or including EED interactions.}\label{table.optimality}
\begin{tabular}{ |c|c|c|c|c|c|c|c|c|c|c|c|c|c|c|  }
 \hline
 & \multicolumn{3}{|c|}{$[\mathrm{FAD}^{\bullet-}, \mathrm{Z}^{\bullet}]$} & \multicolumn{3}{|c|}{$[\mathrm{FAD}^{\bullet-}, \mathrm{Z}^{\bullet}]$, EED} & \multicolumn{4}{|c|}{$[\mathrm{FAD}^{\bullet-}, \mathrm{W}^{\bullet +}]$} &  \multicolumn{4}{|c|}{$[\mathrm{FAD}^{\bullet-}, \mathrm{W}^{\bullet +}]$, EED} \\
 \hline
 $N\pazocal{F}_{\theta} \cdot \Delta^{2}\theta$ & ErC &ErD &AtC& ErC &ErD &AtC & ErC &ErD & ErC/ErD &AtC &ErC &ErD & ErC/ErD &AtC\\
 \hline
 $N_{\hat{I}}$=1   & 10  & 10 &  10& 129   &27&   54 & 10 & 10 & 10 & 10 & 129 & 27 & 51 & 54\\
 \hline
  max &  10  & 10 & 10 & 146   &27&   77 & 11 & 19 & 10 & 10 & 195 & 32 & 132 & 120 \\
 \hline
 min &   4  & 4  & 4  &  90    &16&   39 & 4 & 3 & 4 & 3 & 8 & 11 & 7 & 7 \\
\hline
 avg &   4 &  4  &  4 &  98   &19&   59 & 6 & 5 & 6 & 5 & 17 & 21 & 19 & 21 \\
 \hline
\end{tabular}
\end{table}

For $[\mathrm{FAD}^{\bullet-}, \mathrm{Z}^{\bullet}]$ with EED interactions included, shown in figure \ref{fig.precision_hyperfine}(b), we observe an overall fall in both $\pazocal{F}_{\theta}$ and $1/(N\Delta^{2}\theta)$, with a more pronounced impact on ErC and AtC, due to stronger EED interactions. Nonetheless, the precision of these systems is broadly robust to the inclusion of more nuclei, exhibiting only a slight downwards trend. By analysing $N\pazocal{F} \cdot \Delta^{2}\theta$ a significant overall reduction in the closeness to the QCRB is observed. Varying the number of nuclei in the range $N_{\hat{I}}=1$--$10$, ErC and ErD exhibit an upwards trend with $N_{\hat{I}}$, whilst AtC shows a downwards trend. In figure \ref{fig.heatmaps_hyperfine_1}(b) we have shown representative cases of ErC with $N_{\hat{I}}=10$ and ErD with $N_{\hat{I}}=8$. In general, the product ($N\pazocal{F}_{\theta}\cdot\Delta^{2}\theta)^{-1}$ reveals that the closest approaches towards the QCRB are realised along the axis of maximal $\Phi_{\mathrm{S}}$ contrast, with a more pronounced effect in the case of ErC and ErD. 

Considering $[\mathrm{FAD}^{\bullet-}, \mathrm{W}^{\bullet +}]$ type systems whilst neglecting EED interactions we generally observe a diminished robustness to the number of nuclei, in contrast to the reference-probe systems. Both $\pazocal{F}_{\theta}$ and $1/(N\Delta^{2}\theta)$ values demonstrate a downwards trend as $N_{\hat{I}}$ increases. Similar performance is seen among ErC, ErD, and AtC; however, the composite system ErC/ErD reveals marked resilience to the number of nuclei both in a near order of magnitude increase over its counterparts in $\pazocal{F}_{\theta}$ and $1/(N\Delta^{2}\theta)$, and by plateauing. No clear trend emerges when assessing $N\pazocal{F} \cdot \Delta^{2}\theta$, with averages placing at approximately 10 and below. In figure \ref{fig.heatmaps_hyperfine_2}(a) ErC with $N_{\hat{I}}=8$ and ErC/ErD with $N_{\hat{I}}=10$ are chosen to demonstrate some general patterns shared among AtC, ErC, ErD and ErC/ErD. Namely, the axiality present at $N_{\hat{I}}=1$ no longer holds as $N_{\hat{I}}$ is increased to $N_{\hat{I}}=3$--$4$, but the axis with largest $\Phi_{\mathrm{S}}$ contrast is generally observed in the range of $\phi = 0.25$ to $\phi=0.5$. Along this axis is predominantly where the maximum values of $\pazocal{F}_{\theta}$ and $1/(N\Delta^{2}\theta)$ are found, as well as the closest approaches to the QCRB. An interesting observation arises in the case of AtC and ErD in which, for some choices of $N_{\hat{I}}$, $\pazocal{F}_{\theta}$ remains close to its maximum value across the majority of $\theta$ for a particular range of $\phi$. The composite system also consistently acquires this property for $N_{\hat{I}}>4$ as displayed in figure \ref{fig.heatmaps_hyperfine_2}(a) for $N_{\hat{I}}=10$. 

The inclusion of EED in $[\mathrm{FAD}^{\bullet-}, \mathrm{W}^{\bullet +}]$ results in a further drop in $\pazocal{F}_{\theta}$ and $1/(N\Delta^{2}\theta)$. ErC and AtC systems, constituting those with the largest EED interactions, are most affected. Interestingly, ErC/ErD can occasionally outperform ErD in terms of the $\pazocal{F}_{\theta}$ and come close to the value of $1/(N\Delta^{2}\theta)$ as $N_{\hat{I}}$ increases to 10. Whilst rising trends emerge in $N\pazocal{F}_{\theta}\cdot \Delta^{2}\theta$, which is clear to see from the average against the $N_{\hat{I}}=1$ value in table \ref{table.optimality}, there are some deviations from this trend for the systems with EED. To elucidate the cause of this deviation, we have plotted ErC and ErC/ErD at $N_{\hat{I}}=10$ for $\theta$ and $\phi$ in figure \ref{fig.heatmaps_hyperfine_2}(b). We observe that the largest contrast in $\Phi_{\mathrm{S}}$ might not lie directly along $\theta$. In cases where the axis of the largest contrast occurs predominantly along $\phi$, this can lead to a significant decrease in $1/(N\Delta^{2}\theta)$, as it partly relies on $\mathrm{d}\Phi_{\mathrm{S}}/\mathrm{d}\theta$. Conversely, $\pazocal{F}_{\theta}$ is not as restricted, showing that there are alternative measures capable of estimating $\theta$ more precisely in these circumstances. This is more pronounced in the systems with larger EED interaction and large $N_{\hat{I}}$, namely, ErC and AtC, with $N_{\hat{I}}=9$ and $N_{\hat{I}}=10$.

Drawing together observations we find, in general, that the magnitude of $\pazocal{F}_{\theta}$ and $1/(N\Delta^{2}\theta)$ is larger in $[\mathrm{FAD}^{\bullet-}, \mathrm{Z}^{\bullet}]$. This system is more robust to the inclusion of more coupled nuclei and, in some cases, exhibits a clear trend approaching the QCRB. However, $[\mathrm{FAD}^{\bullet-}, \mathrm{W}^{\bullet +}]$ can, at specific $N_{\hat{I}}$ and on average, approach optimality more than the reference-probe system. This becomes even more evident with the inclusion of EED interactions, where every $[\mathrm{FAD}^{\bullet-}, \mathrm{W}^{\bullet +}]$ system finds several values of $N_{\hat{I}}$ at which it can outperform its reference-probe counterpart, approaching the QCRB more closely. This is also true of the averages for ErC, AtC. Additionally, the composite system can provide a boost over ErC exhibiting a larger magnitude of $\pazocal{F}_{\theta}$ and $1/(N\Delta^{2}\theta)$, and showing robustness through plateauing as $N_{\hat{I}}$ is increased. In the SM heatmaps, we demonstrate that even closer approaches to the QCRB are possible over $\theta$ and $\phi$, though this can occur at correspondingly low values of $1/(N\Delta^{2}\theta)$ and $\pazocal{F}_{\theta}$. Overall, whilst some trends emerge in the approach to optimality as $N_{\hat{I}}$ is increased, and on average, we observe the more complex systems comprising many coupled nuclei are closer to optimality than their counterparts comprising only a single nuclei, there is still room to saturate the QCRB and improve the measurement precision.  

\section{Discussion}
We aimed to quantify the precision of the radical-pair compass and evaluate how optimal it is in the limits of biological complexity, characterised by a large number of hyperfine coupled nuclei and presence of electron-electron dipolar coupling. Utilising the quantum Fisher information to set a bound on the ultimate limits of precision, via the quantum Cram\'er--Rao bound, we examined cryptochromes ErC, ErD, and AtC for reference-probe, $[\mathrm{FAD}^{\bullet-}, \mathrm{Z}^{\bullet}]$, and FAD/tryptophan, $[\mathrm{FAD}^{\bullet-}, \mathrm{W}^{\bullet +}]$, radical-pair models and a composite system ErC/ErD for $[\mathrm{FAD}^{\bullet-}, \mathrm{W}^{\bullet +}]$. Our findings indicate that, an increase in the number of hyperfine coupled nuclei can lead to trends towards optimal precision in yield-based measures as shown in table \ref{table.optimality} by a decrease in $N\pazocal{F}_{\theta} \cdot \Delta^{2}\theta$ from the value at $N_{\hat{I}}=1$ to the minimum and average values. However, in general the achieved precision falls short of optimality, i.e.\ saturating the QCRB, by 1-2 orders of magnitude, suggesting scope for improvement. 

The radical-pair mechanism as currently envisioned provides a qualitative explanation of characteristic traits of magnetoreception. Yet, conceptual challenges persist, such as the sensitivity gap between ethological observations and predictions based off simulations of realistically complex models. Specifically, model studies have shown that, as the number of hyperfine interactions is increased to approach the number inherent in the biological system, the sensitivity is significantly compromised, questioning the viability of the mechanism \cite{Atkins2019, Babcock2020}. Furthermore, it has been raised that unavoidable inter-radical interactions, in particular the EED interaction, further limit the attainable precision \cite{Babcock2020}. Investigations have responded to these challenges by attempting to identify additional or alternative mechanisms that can restore and boost sensitivity despite the presence of detrimental interactions. The RPM as currently established, is still the benchmark model and thus investigated here, but it might be interesting to explore phenomena that promises larger magnetic field effects, allowing the optimal limit of precision set by quantum mechanics to be more closely approached. For example, radical triads \cite{Babcock2021, Kattnig2017, Kattnig2017a}, the quantum Zeno effect \cite{Ramsay2022, Dellis2012, Kominis2009}, the chiral induced spin selectivity (CISS) effect \cite{Poonia2023, Poonia2022, Rahman2021}, and radical motion\cite{Ramsay2023, Smith2022a, Kattnig2016b} have all been shown to provide enhancements to sensitivity, and possibly restore compass function. 

In $[\mathrm{FAD}^{\bullet-}, \mathrm{Z}^{\bullet}]$ we observed the largest magnitude and robustness in precision and compass contrast as assessed by the yield anisotropy, as the number of nuclei increased, with more pronounced trends towards optimal precision when neglecting EED. However, in the presence of EED interactions, which are unavoidable in immobilised radical-pairs and cause an overall drop in magnitude of compass precision and contrast, we found that $[\mathrm{FAD}^{\bullet-}, \mathrm{W}^{\bullet +}]$ can approach the QCRB more closely, as shown by the minimum values of $N\pazocal{F}_{\theta} \cdot \Delta^{2}\theta$ in table \ref{table.optimality}, than reference-probe systems with the same number of nuclei and, on average, are more optimal. Nevertheless, the $[\mathrm{FAD}^{\bullet-}, \mathrm{W}^{\bullet +}]$ with EED interactions exhibit more complex profiles in magnetic field parameters $\theta$ and $\phi$ due to a loss of axiality, which can lead to a sub-optimal scanning axis when assessing $N\Delta^{2}\theta$ in specific examples and occasional spikes in $N\pazocal{F}_{\theta} \cdot \Delta^{2}\theta$ to low optimality ratios. Remedying this would require a multiparameter estimation study and calculation of the quantum Fisher information matrix \cite{Ji2019}. Such a study would constitute significant computational challenge and is beyond the scope of this study. However, as such an approach would produce smoother $N\pazocal{F}_{\theta} \cdot \Delta^{2}\theta$ curves for $[\mathrm{FAD}^{\bullet-}, \mathrm{W}^{\bullet +}]$, it is clear that it would corroborate the conclusion that $[\mathrm{FAD}^{\bullet-}, \mathrm{W}^{\bullet +}]$ systems approach the quantum limit closer than $[\mathrm{FAD}^{\bullet-}, \mathrm{Z}^{\bullet}]$.

The greater rise in optimality observed in more complex models of $[\mathrm{FAD}^{\bullet-}, \mathrm{W}^{\bullet +}]$, i.e with EED interactions and a larger number of nuclei, in comparison to the more idealised and simpler $[\mathrm{FAD}^{\bullet-}, \mathrm{Z}^{\bullet}]$, points to a remarkable possibility that nature may have evolutionary adapted hyperfine interactions and the inter-radical axis to optimise precision. Indeed, this optimisation can even, in some instances, surpass models of the $[\mathrm{FAD}^{\bullet-}, \mathrm{Z}^{\bullet}]$ radical-pair unrealistically neglecting EED interactions. Our observations indicate that systems with larger EED couplings, exhibit a more significant rise in optimality associated with an increase in number of nuclei. This could indicate that low effect sizes have served as drivers of an optimisation that led to, or necessitated, a closer approach to the QCRB. However, care must be taken in interpreting trends along with evolutionary adaptations, as it is unlikely that nature could finely  control the number of hyperfine interactions as we have done to probe the limits of biological complexity. Instead, it is plausible that cryptochrome evolved from a common ancestor with photolyases, developing magnetic field sensitivity in an already complex biological setting \cite{Bartolke2021, Ozturk2017, Sancar2003, Liedvogel2010, Cashmore1999}, for which numerous hyperfine couplings and EED interactions would be relevant. As such, it would be of interest for future studies to consider the radical-pair precision and contrast following the evolutionary chain and adaptations of cryptochromes as well as in variants with different function and structures to understand the evolutionary optimisation of magnetoreception, discernible as an approach towards the QCRB.  

One potential advantageous adaptation may have occurred in cryptochromes possessing the $\mathrm{W_{D}}$ tryptophan in the form of a fast electron transfer between $\mathrm{W_{C}}$ and $\mathrm{W_{D}}$, thereby incorporating both within a composite radical-pair. This is corroborated by predicted rate constants of degenerate electron exchange in a cryptochrome of a migratory bird, as provided by \cite{Xu2021}. It has been proposed that this may play a role in ensuring that both the sensing and signalling capabilities of cryptochrome are satisfied, where $\mathrm{W_{D}}$ was considered relevant to the signalling process \cite{Xu2021}. Here we have observed that this system not only facilitates larger magnitudes of precision, but also  exhibits greater robustness to an increase in hyperfine interactions and the presence of EED interactions, where in the latter case it benefits from the increased inter-radical distance of $\mathrm{W_{D}}$. This observation might, in part, offer a potential explanation for why animal cryptochromes possess the $\mathrm{W_{D}}$ tryptophan, i.e.\ to provide it superior precision. Whilst the ErD model also exhibits some of these properties due to its reduced EED interaction, it should be noted that it is unlikely to be realised independently due to insufficiently fast recombination reaction kinetics, which at the larger distance would not permit out-competing the estimated spin relaxation rates, thus implying loss of compass sensitivity. Lastly, our findings reveal that the composite system in the limits of an increasing number of hyperfine interactions acquires a consistent precision, as assessed by the QFI, along the entire range of $\theta$ values, suggesting there are radical-pair design principles and measurements that may provide robust precision for a quantum compass.  

On the other hand, one should not disregard the reference-probe model $[\mathrm{FAD}^{\bullet-}, \mathrm{Z}^{\bullet}]$ as its larger magnitude contrast and precision, together with clear trend behaviour in $N\pazocal{F}_{\theta} \cdot \Delta^{2}\theta$ as number of nuclei is increased, make it desirable. Furthermore, here we have assessed it in comparison to the $[\mathrm{FAD}^{\bullet-}, \mathrm{W}^{\bullet +}]$ and have as such chosen inter-radical distances, EED tensors, and radical lifetimes in accordance with this. Therefore, our findings do not exclude the possibility that an alternative selection of parameters and mechanisms might be more conducive. Radical-pairs where this might be the case include the flavin semiquinone/superoxide radical-pair $[\mathrm{FADH}^{\bullet}/ \mathrm{O_{2}}^{\bullet -}]$ radical-pair, though the fast spin relaxation the radical-pair is subject to poses challenges to provide a sufficient coherence lifetime to permit magnetic field sensitivity. 

One might anticipate, \textit{a priori}, that the geometric closeness of the measurement operator to the optimal estimator might correlate with rising trends to optimal precision. However, we observed that yield measurements are, in general, close to orthogonal to the optimal estimator and tend closer towards orthogonality as the number of nuclei is increased as shown in Appendix \ref{sec.appendix_3}. This implies that the similarity of the measurement operator to the optimal estimator itself is insufficient to predict its precision capabilities. Moreover, it suggests that precision in radical-pair systems is not confined to a specific measurement operator, and that the optimal limit of precision can still be approached regardless of lack of similarity to the optimal estimator. Even restricted to mere yields, we have observed that the radical-pair compass can approach optimality in the limits of biological complexity. Furthermore, we have found that the Cram\'er-Rao bound set by $F_{\theta}$ is saturated, thus an increase in the precision attainable, up to the QCRB would only be possible if a different measurement was possible. An example of this might be found in a measurement where the measurement elements are associated with the eigenstates of $\hat{S}^{2}$, thereby resolving the singlet state and individual triplet states. Future investigations could explore the measurements accessible under laboratory conditions to determine the feasibility of the optimal estimator. If it is unattainable, as expected, the focus could shift to understanding how much more precisely we could measure weak magnetic fields using radical-pairs compared to nature.  

Whilst we see that the QCRB can be approached in more complex systems, i.e.\ by including a larger number of hyperfine interactions and EED interactions, a gap still remains to saturating it. To some extent this could be because the measure that optimises precision in radical-pairs may not be accessible to nature, as it must work within biological constraints and with the objective to elicit a biologically interpretable sensory response, where the consensus view is that reaction yields are pertinent. In relation to this we have observed that the largest yield-based precision in $\theta$ tends to occur along an axis corresponding to maximal contrast in singlet yields, particularly at points characterised by the sharpest change, i.e.\ relating to $\partial_{\theta}\Phi_{\mathrm{S}}$. Nonetheless, the question remains as to whether the remaining gap can be closed by any additional mechanisms. Previously, we have shown that periodically driven radical motion can enhance the anisotropy and coherence \cite{Smith2022a}, and is also capable of enhancing precision in the context of thermometry \cite{Glatthard2022}. It would thus be prudent to question, and is the focus of an upcoming study, whether driving of this form could also increase the precision capabilities of the radical-pair, help to approach the QCRB, and whether this comes in addition to the enhancement to anisotropy, or as a compromise. Similarly, the QFI could be used to analyse the advantages of other proposed mechanisms that have demonstrated increased sensitivity in the face of inter-radical interactions, such as three-radical mechanisms, the CISS effect, the quantum Zeno effect, and radical motion arising from environment interaction. This would elucidate relations between compass precision, contrast, and coherence \cite{Kominis2023}, environment parameters \cite{Mirza2023}, and viability in nature and quantum technologies \cite{Paris2011}.

\section{Conclusions}

Quantum metrology establishes fundamental bounds on the precision with which a parameter can be inferred through measurements on a quantum system. Avian magnetoreception and several related magnetosensitive traits of various animals are thought to rely on spin dynamics in a protein, suggesting the existence of a quantum magnetometer in biology. While this biological magneto-sensor utilises quantum phenomena, it must simultaneously permit a biological readout, which severely constrains the kind of measurements that can be realised. In fact, only spin-selective recombination reactions of the radical-pair are widely considered innate. This raises the question as to what extent such constrained systems approach the precision bounds of quantum measurements as set by the quantum Cram\'er--Rao bound. Here, we find that a yield-based measurement in general falls short of the optimal measurement by factors of the order of  $10$--$100$. However, we also observe that, by systematically increasing the complexity of the radical-pair through incorporating successively more hyperfine interactions, more complex systems exhibit a closer approach towards the optimal precision possible. This is particularly evident when the oft-neglected, but unavoidable EED interaction is properly included. Specifically, our systematic simulations suggested that FAD/tryptophan ($[\mathrm{FAD}^{\bullet-}, \mathrm{W}^{\bullet +}]$) radical-pairs have evolved to a higher degree, which might be a result of inherently smaller magnetic field effects posing significant evolutionary pressure. A ErC/ErD radical-pair involving fast degenerate electron exchange between tryptophan sites ($[\mathrm{FAD}^{\bullet-}, \mathrm{W_{C}}^{\bullet +}/\mathrm{W_{D}}^{\bullet +}]$), practically feasible (as established in \textit{in vitro} experiments),  appears to be well optimised in terms of approaching the quantum measurement limit and comparably effective when considering magnitudes of compass contrast and precision. The better separated ErD model ($[\mathrm{FAD}^{\bullet-},\mathrm{W_{D}}^{\bullet +}]$) might be preferable in view of delivered compass sensitivity, due to its comparably reduced EED interactions, but might not be realisable due to a recombination rate that is too small as a consequence of the large separation of FAD and $\mathrm{W_{D}}$. We observed the system that attained the closest precision was the $[\mathrm{FAD}^{\bullet-}, \mathrm{W}^{\bullet +}]$ AtC radical-pair with $N_{\hat{I}} = 6$, however, the $[\mathrm{FAD}^{\bullet-}, \mathrm{W}^{\bullet +}]$ ErD radical-pair performed best when including EED interactions and with $N_{\hat{I}}=10$. The best systems when assessing models by average performance over $N_{\hat{I}}$, and when considering the more realistic treatment of including EED interactions, were ErC, closely followed by ErC/ErD. FAD/Z-systems ($[\mathrm{FAD}^{\bullet-}, \mathrm{Z}^{\bullet}]$) generally give rise to larger compass contrast, while they overall appear to be less optimised. However, in the ongoing absence of graspable insights in the identity of Z, the practicality of the model remains questionable despite its popularity. In principle, there is room for improvement for all studied radical-pair systems, potentially realisable by going beyond yield measurements. However, we have identified that the optimal measures do not typically resemble a singlet state measurement, and it is not obvious how alternative measures could be transduced into a chemical signal. It appears that the constraints set in place by radical chemistry and biology are strong and paramount over quantum optimality.

Sensitivity to magnetic fields is ubiquitous in nature and radical mechanisms potentially underpin several biological processes \cite{Krylov2023,Zadeh-Haghighi2022}, such as neurogensis \cite{Ramsay2022}, magnetic field and lithium effects on the circadian clock \cite{Zadeh-Haghighi2022a}, and processes involving reactive oxygen species \cite{Rishabh2022}, thereby forming an area of intense interest to the field of spin chemistry \cite{Hore2020} and a cornerstone of quantum biology \cite{Kim2021, Marais2018}. Here we have demonstrated that the quantum Fisher information, and the quantum Cram\'er-Rao bound it sets on precision, is an effective tool, in addition to the conventional contrast measure, to assess fundamental limits of precision of different radical models against one another in the limits of biological complexity and identify opportunities to improve upon natures design principles. Furthermore, it provides scope for future studies to analyse additional mechanisms, such as driven radical motion, and elucidate those that provide the optimal enhancements in precision, thereby giving insight into how nature may achieve exquisite magnetic field sensitivity and how we can optimise it for technological applications.  


\titleformat{\section}
  {\normalfont\bfseries\centering}{\thesection}{1em}{}
\renewcommand*{\thesection}{}
\section{Acknowledgements}

We acknowledge use of the University of Exeter's HPC facility. This work was supported by the Office of Naval Research Global (ONR-G award number N62909-21-1-2018), the Leverhulme Trust (RPG-2020-261), and the Engineering and Physical Sciences Research Council (EPSRC grants EP/V047175/1 and EP/X027376/1). We thank Prof.\ Ilia Solov'yov and Gesa Gr\"{u}ning (University of Oldenburg, Germany) for providing protein structural data and raw hyperfine parameters as derived in reference \cite{Gesa2022}.

\section*{Data availability}
The data that support the findings of this study are available from the corresponding author upon reasonable request.
\appendix
\section{Derivation of variance from error propagation} \label{sec.appendix_1}
In the main text we have derived $\Delta^{2}\theta$ associated with measuring recombination yields by considering the recombination success as binomially distributed. More generally, the error associated with measuring observable $\hat{O}$ through $N$ independent measurements is given by 
\begin{align}
    \Delta^{2}\theta &= \frac{\Delta^{2}\hat{O}}{N\vert \frac{\mathrm{d}\langle \hat{O} \rangle}{\mathrm{d}\theta}\vert^{2}}. \label{eq.error_prop}
\end{align}
It is curious to note that if one uses the general propagation of error formula with the square of the total angular momentum as the observable $\hat{O}=\langle \hat{S}^{2} \rangle $, and using the property that
\begin{align}
\hat{S}^{2} = (\hat{S}_{\mathrm{A}} + \hat{S}_{\mathrm{B}})^{2} = 3/2 + 2\hat{S}_{\mathrm{A}} \cdot \hat{S}_{\mathrm{B}} = 2(\pmb{1}-\hat{P}_{\mathrm{S}}),    
\end{align}
as $\hat{P}_{\mathrm{S}} = \pmb{1}/4 - \hat{S}_{\mathrm{A}}\cdot \hat{S}_{\mathrm{B}}$, we obtain
\begin{align}
    \Delta^{2}\theta = \frac{\Delta^{2}\hat{S}^{2}}{N\vert\mathrm{\frac{d\langle \hat{S}^{2} \rangle}{\mathrm{d}\theta}\vert^{2}}} 
    = \frac{\langle \hat{S}^{4} \rangle  - \langle \hat{S}^{2} \rangle^{2}} {N\vert\mathrm{\frac{d\langle \hat{S}^{2} \rangle}{\mathrm{d}\theta}\vert^{2}}} 
     = \frac{4(\pmb{1}-\Phi_{\mathrm{S}}) - 4(\pmb{1}-\Phi_{\mathrm{S}})^{2}}{N\vert \frac{\mathrm{d}}{\mathrm{d}\theta} 2(\pmb{1}-\Phi_{\mathrm{S}}) \vert^{2}} 
    =\frac{\Phi_{\mathrm{S}}(\pmb{1}-\Phi_{\mathrm{S}})}{N\vert\mathrm{\frac{d\Phi_{\mathrm{S}}}{\mathrm{d}\theta}\vert^{2}}},
\end{align}
which equates to the result of equation \ref{eq.variance} as in \cite{Guo2017}. 
\section{Relationship between CFI and variance} 
\label{sec.appendix_2}
We show that $\Delta^{2}\theta$ saturates the Cram\'er-Rao bound set by $F_{\theta}$ with measurement elements $\hat{\Pi}_{n} \in \{ \hat{P}_{\mathrm{S}}, \hat{P}_{\mathrm{T}} \}$. Using the definition of the classical Fisher information in equation \ref{eq.defcfi} we obtain
\begin{align}
    F_{\theta} &= \frac{(\partial_{\theta}\mathrm{Tr}[\hat{P}_{\mathrm{S}}\hat{\rho}_{\theta}])^{2}}{\mathrm{Tr}[\hat{P}_{\mathrm{S}}\hat{\rho}_{\theta}]} + \frac{(\partial_{\theta}\mathrm{Tr}[(\pmb{1} - \hat{P}_{\mathrm{S}})\hat{\rho}_{\theta}])^{2}}{\mathrm{Tr}[(\pmb{1} - \hat{P}_{\mathrm{S}})\hat{\rho}_{\theta}]} \nonumber \\
    &= \frac{(\partial_{\theta} \Phi_{\mathrm{S}})^{2}}{\Phi_{\mathrm{S}}} + \frac{(\partial_{\theta}\Phi_{\mathrm{S}})^{2}}{1-\Phi_{\mathrm{S}}} \nonumber \\
    &= \frac{(\partial_{\theta} \Phi_{\mathrm{S}})^{2}}{\Phi_{\mathrm{S}}(1-\Phi_{\mathrm{S}})}\nonumber \\
    &= \frac{1}{N\Delta^{2}\theta},
\end{align}
where $\Phi_{\mathrm{S}} = \mathrm{Tr}[\hat{P}_{\mathrm{S}}\hat{\rho}_{\theta}]$ is the singlet yield. Thus, the Cram\'er-Rao bound set by $F_{\theta}$ is saturated, meaning the information is fully utilised, and additional precision is only possible through consideration of different measurements, up to the limits of the quantum Cram\'er-Rao bound set by $\mathcal{F}_{\theta}$. This is expected as, in general, the Cram\'er-Rao bound of the variance of the mean number of successes of a single-parameter Bernoulli trial is an equality.

\section{Orthogonality distance to optimal estimator} \label{sec.appendix_3}
\begin{figure}
\centering
	\includegraphics[scale=1]{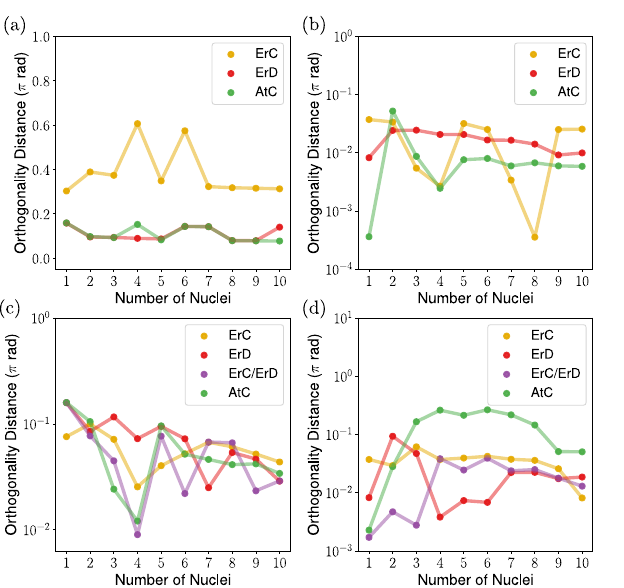}
	\caption{The distance from orthogonality between $\hat{S}^{2}$ and the optimal estimator for varying number of nuclei. Here, we use equivalent models to figure \ref{fig.anisotropy}. Reference-probe $[\mathrm{FAD}^{\bullet-}, \mathrm{Z}^{\bullet}]$ radical-pairs are considered without EED in (a) and including EED interactions in (b). FAD/tryptophan $[\mathrm{FAD}^{\bullet-}, \mathrm{W}^{\bullet +}]$ radical-pairs are considered without EED in (c) and including EED interactions in (d).}
    \label{fig.orthog_dist}
\end{figure}
By constructing the optimal estimators $\hat{M}_\theta$ using equation \ref{eq.opt_est} we assess the similarity between $\hat{S}^{2}$ and $\hat{M}_{\theta}$. Specifically, we decompose the operators in terms of the 16 spin components $\hat{\mathcal{S}}_{i}$ 
 $\in$ $\{\frac{1}{2} \pmb{1}\pmb{1}, \pmb{1}$ $\hat{S}_{x}$, \pmb{1}$\hat{S}_{y}$, \pmb{1}$\hat{S}_{z}$, ..., $2\hat{S}_{z}$$\hat{S}_{z}$\}, which forms a basis of the electronic Liouville space, orthonormal with respect to the Hilbert-Schmidt inner product. The expansion coefficients are then evaluated as $\mathbf{c}_{i} = \mathrm{Tr}[\hat{O}\hat{\mathcal{S}}_{i}]$. Thus, by constructing all coefficients for $\hat{M}_{\theta}$, and $\hat{O}$, denoted $\mathbf{c}_{\hat{O}}$ and $\mathbf{c}_{\hat{M}_{\theta}}$ respectively, we assess the similarity of the operators via the angle between spin components
\begin{align}
    \alpha = \mathrm{arccos}\bigg(\frac{\mathbf{c}_{\hat{M}_{\theta}}\cdot \mathbf{c}_{\hat{O}}}{\vert \mathbf{c}_{\hat{M}_{\theta}} \vert \vert \mathbf{c}_{\hat{O}} \vert}\bigg).
\end{align}

As the vast majority of our results show that $\hat{M}_{\theta}$ and $\hat{O}$, here chosen as $\hat{S}^{2}$, are close to orthogonal, we opt to measure the distance from orthogonality, i.e.\ $\vert \alpha - \pi/2 \vert $. The results are shown in figure \ref{fig.orthog_dist}, for the discussed radical-pair models, $[\mathrm{FAD}^{\bullet-}, \mathrm{Z}^{\bullet}]$ and $[\mathrm{FAD}^{\bullet-}, \mathrm{W}^{\bullet +}]$. As stated, the majority of results are close to orthogonal, which in part is due to the high dimensionality of the two-spin spin components, meaning that there are many ways to be close to orthogonal and less ways for the operators to be similar. However, a further trend towards orthogonality is observed as the number of nuclei is increased, and with the inclusion of EED interactions. In the SM (figures S22-S25) we also analyze the orthogonality distance between the optimal estimator and alternative measurements, $\hat{P}_{\mathrm{S}}$, $\hat{S}_{x}$, $\hat{S}_{y}$ and $\hat{S}_{z}$, where similar observations arise. 
\begin{figure}[t!]
\centering
	\includegraphics[scale=1]{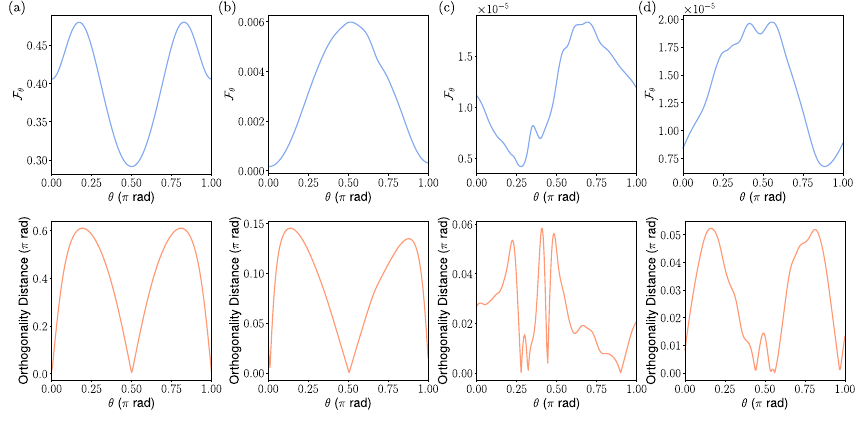}
	\caption{$\pazocal{F}_{\theta}$, and the distance from orthogonality between $\hat{S}^{2}$ and the optimal estimator for select cases of radical-pairs with $N_{\hat{I}}=10$ nuclei against $\theta$. Specifically, we consider reference-probe $[\mathrm{FAD}^{\bullet-}, \mathrm{Z}^{\bullet}]$ radical-pairs are shown in (a) for ErC neglecting EED interactions and in (b) ErC including EED interactions. $[\mathrm{FAD}^{\bullet-}, \mathrm{W}^{\bullet +}]$ radical-pairs are shown in (c) for ErC including EED interactions and in (d) for AtC including EED interactions.}
    \label{fig.orthg_dist_2}
\end{figure}

In figure \ref{fig.orthg_dist_2} we present $\pazocal{F}_{\theta}$, and the orthogonality distance resolved over $\theta$ for select examples of $[\mathrm{FAD}^{\bullet-}, \mathrm{Z}^{\bullet}]$ and $[\mathrm{FAD}^{\bullet-}, \mathrm{W}^{\bullet +}]$ radical-pairs with $N_{\hat{I}}=10$ nuclei. For a ErC $[\mathrm{FAD}^{\bullet-}, \mathrm{Z}^{\bullet}]$ radical-pair neglecting EED interactions, it is observed that the orthogonality distance increases are also accompanied by an increase in $\pazocal{F}_{\theta}$. However, this relationship is lost upon inclusion of EED, and instead a stronger correlation of the orthogonality distance and the gradient of $\pazocal{F}_{\theta}$ is observed. Although similar observations arise in $[\mathrm{FAD}^{\bullet-}, \mathrm{W}^{\bullet +}]$ radical-pairs when including EED interactions, overall the trend is less pronounced. Nevertheless, it is clear that the measure of $\hat{S}^{2}$ is more auspicious for certain values of $\theta$. 

\bibliography{qfi_ref_tidy}

\end{document}